# Analytic $C_{\ell_1}$ Norm of Coherence Evolution for Bell States under a Two-Qubit Superconducting Hamiltonian

Seyed Mohsen Moosavi Khansari[1]
*Department of Physics, Faculty of Basic Sciences, Ayatollah Boroujerdi University, Boroujerd, IRAN*

**Abstract**

We present the first exact analytic study of unitary coherence dynamics in a minimal two-qubit superconducting system with $zz$-coupling and transverse Josephson tunneling. By deriving the full time-evolution operator without approximations and propagating all four Bell-state initial conditions, we obtain novel closed-form time-dependent pure-state density matrices and an explicit analytic expression for the $C_{\ell_1}$ norm of coherence that, to our knowledge, has not been previously reported in the literature. Two of the Bell states are shown to be invariant under the model dynamics with constant coherence, whereas the other two exhibit controlled, parameter-dependent coherence oscillations. The oscillatory behavior is governed by two distinct frequency scales that directly map onto the circuit coupling and tunneling parameters, allowing for predictable tuning of amplitude and periodicity. Numerical visualizations clarify operating regimes for transient enhancement or suppression of coherence. These results deliver compact, analytically tractable tools for parameter optimization and provide a clear foundation for incorporating dissipation and experimental validation.

**Keywords:** Quantum coherence, $C_{\ell_1}$ norm of coherence, Bell states, Superconducting qubits, Coherence control

## 1. Introduction

Quantum coherence and entanglement are central resources for quantum information processing, underpinning protocols in computation, communication, and metrology. Superconducting qubits have emerged as a leading platform for scalable quantum processors due to their lithographic manufacturability, fast gate speeds, and steadily improving coherence times. In coupled superconducting architectures, the interplay between inter-qubit coupling and single-qubit tunneling determines both the generation and temporal evolution of quantum correlations; therefore, analytic control of these dynamics is crucial for device design and operation [1–8].

In this work, we present an exact, closed-form analysis of unitary coherence dynamics in a minimal two-qubit superconducting system (TQS) governed by a Hamiltonian comprising a $zz$-type mutual coupling energy $E_m$ and transverse Josephson tunneling $E_J$. The novelty of our approach lies in: () the derivation of the **exact** $4 \times 4$ unitary propagator without employing rotating-wave or perturbative approximations, (ii) the analytic propagation of **all four Bell states** to yield explicit density matrices (the density matrix $\rho(t)$ is a positive semi-definite Hermitian operator that completely describes the quantum state, with diagonal elements representing populations and off-diagonal elements representing coherences), (iii) the first closed-form expression for the $C_{\ell_1}$ norm of coherence for this system, (iv) the rigorous proof that $|\Phi^-\rangle$ and $|\Psi^-\rangle$ are eigenstates with constant coherence, and (v) the identification of two distinct frequency scales that enable the predictable tuning of coherence oscillations. These results establish a comprehensive analytic framework that complements and extends prior numerical and approximate studies.

[1] M.Moosavikhansari@abru.ac.ir

By propagating Bell state initial conditions under the derived $4 \times 4$ time evolution operator $U$, we obtain explicit time-dependent pure state density matrices and compute the $C_{\ell_1}$ norm of coherence analytically. This approach reveals which Bell states are invariant under the model dynamics and which exhibit parametrically tunable coherence oscillations, identifying the relevant frequency scales and their dependence on $E_J$ and $E_m$. Complementary numerical visualizations clarify amplitude and periodicity regimes and demonstrate how circuit parameters can be used to engineer transient or steady coherence resources. Our closed-form results provide compact formulas suitable for rapid parameter scans and optimization, and establish a baseline for future inclusion of dissipation, noise, and experimental validation via state tomography (quantum state tomography (QST) is an experimental procedure that reconstructs the density matrix of a quantum state through a series of projective measurements).

## 2. Theoretical Framework: Hamiltonian and Time-Evolution Operator

### 2.1 Hamiltonian of the Two-Qubit Superconducting System

The Hamiltonian of a Two-Qubit Superconducting system (TQS), consisting of qubits $a$ and $b$, is expressed as

$$\mathcal{H}_{TQS} = \frac{1}{4}\hbar^2 E_m\, \sigma_a^z \otimes \sigma_b^z \; - \; \frac{1}{2}\hbar E_J\, \sigma_a^x \otimes I_b \; - \; \frac{1}{2}\hbar E_J\, I_a \otimes \sigma_b^x \qquad (1)$$

where $E_m$ denotes the mutual coupling energy, and $E_J$ represents the Josephson energy. The operators $\sigma_a^z$ and $\sigma_a^x$ are the Pauli matrices acting on qubit $a$ in the $z$ and $x$ directions, respectively, and similarly $\sigma_b^z$, $\sigma_b^x$ correspond to qubit $b$. Figure 1 provides a detailed illustration of the two-qubit superconducting system (TQS) [9, 10]. The Hamiltonian parameters in Equation (1) correspond to specific experimental quantities: $E_J$ represents the Josephson energy of each transmon junction (typically $E_J/h \sim$ 10–20 GHz), while $E_m$ denotes the mutual coupling energy between qubits, which for capacitive coupling scales as $E_m \propto g$ where $g/2\pi \sim$ 5–50 MHz is the qubit-qubit coupling rate. This mapping enables direct experimental implementation on existing superconducting quantum processors, such as IBM Quantum's fixed-frequency transmons or Google's tunable-coupler devices.

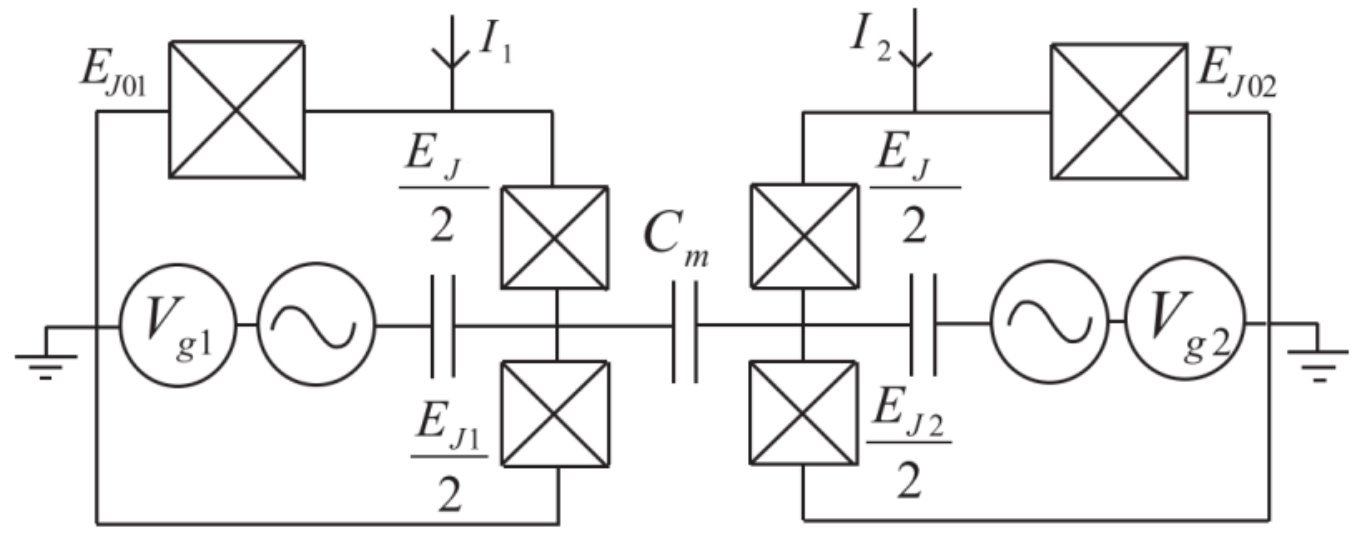

Figure 1: A comprehensive depiction of a two-qubit superconducting system (TQS).

### 2.2 Matrix Form of the Hamiltonian

The Hamiltonian matrix in the computational basis $\{|\,00\rangle, |\,01\rangle, |\,10\rangle, |\,11\rangle\}$ (computational basis $\{|\,00\rangle, |\,01\rangle, |\,10\rangle, |\,11\rangle\}$ where $|\,0\rangle$ and $|\,1\rangle$ denote the ground and excited states of each qubit) is:

$$\mathcal{H}_{TQS} = \begin{pmatrix} \frac{\hbar^2 E_m}{4} & -\frac{\hbar E_J}{2} & -\frac{\hbar E_J}{2} & 0 \\ -\frac{\hbar E_J}{2} & -\frac{1}{4}\hbar^2 E_m & 0 & -\frac{\hbar E_J}{2} \\ -\frac{\hbar E_J}{2} & 0 & -\frac{1}{4}\hbar^2 E_m & -\frac{\hbar E_J}{2} \\ 0 & -\frac{\hbar E_J}{2} & -\frac{\hbar E_J}{2} & \frac{\hbar^2 E_m}{4} \end{pmatrix} \qquad (2)$$

### 2.3 Time-Evolution Operator *U*

Using this Hamiltonian matrix, the unitary time-evolution operator for the TQS system is obtained as:

$$U = \exp\left(-i\,\mathcal{H}_{TQS}\, t/\hbar\right) \qquad (3)$$

$U$ is a $4 \times 4$ matrix whose elements are defined as follows:

$$\begin{aligned} U_{1,1} = U_{4,4} \quad &= -\frac{i\hbar E_m \sin\left(\frac{1}{4}t\sqrt{16E_J^2 + \hbar^2 E_m^2}\right)}{2\sqrt{16E_J^2 + \hbar^2 E_m^2}} + \frac{1}{2}\cos\left(\frac{1}{4}t\sqrt{16E_J^2 + \hbar^2 E_m^2}\right) \\ &\quad -\frac{1}{2}i\sin\left(\frac{1}{4}t\hbar E_m\right) + \frac{1}{2}\cos\left(\frac{1}{4}t\hbar E_m\right) \end{aligned} \qquad (4)$$

$$U_{1,2} = U_{1,3} = U_{2,1} = U_{2,4} = U_{3,1} = U_{3,4} = U_{4,2} = U_{4,3} = \frac{2iE_J \sin\left(\frac{1}{4}t\sqrt{16E_J^2 + \hbar^2 E_m^2}\right)}{\sqrt{16E_J^2 + \hbar^2 E_m^2}} \qquad (5)$$

$$U_{1,4} = U_{4,1} = -\frac{i\hbar E_m \sin\left(\frac{1}{4}t\sqrt{16E_J^2 + \hbar^2 E_m^2}\right)}{2\sqrt{16E_J^2 + \hbar^2 E_m^2}} + \frac{1}{2}\cos\left(\frac{1}{4}t\sqrt{16E_J^2 + \hbar^2 E_m^2}\right) + \frac{1}{2}i\sin\left(\frac{1}{4}t\hbar E_m\right) - \frac{1}{2}\cos\left(\frac{1}{4}t\hbar E_m\right) \tag{6}$$

$$U_{2,2} = U_{3,3} = \frac{i\hbar E_m \sin\left(\frac{1}{4}t\sqrt{16E_J^2 + \hbar^2 E_m^2}\right)}{2\sqrt{16E_J^2 + \hbar^2 E_m^2}} + \frac{1}{2}\cos\left(\frac{1}{4}t\sqrt{16E_J^2 + \hbar^2 E_m^2}\right) + \frac{1}{2}i\sin\left(\frac{1}{4}t\hbar E_m\right) + \frac{1}{2}\cos\left(\frac{1}{4}t\hbar E_m\right) \tag{7}$$

$$U_{2,3} = U_{3,2} = \frac{i\hbar E_m \sin\left(\frac{1}{4}t\sqrt{16E_J^2 + \hbar^2 E_m^2}\right)}{2\sqrt{16E_J^2 + \hbar^2 E_m^2}} + \frac{1}{2}\cos\left(\frac{1}{4}t\sqrt{16E_J^2 + \hbar^2 E_m^2}\right) - \frac{1}{2}i\sin\left(\frac{1}{4}t\hbar E_m\right) - \frac{1}{2}\cos\left(\frac{1}{4}t\hbar E_m\right) \tag{8}$$

### 2.4 Computational Implementation with Python

The analytical solutions derived in this work can be readily implemented numerically using Python, which provides a flexible and accessible environment for simulating and analyzing the two-qubit superconducting system. Python enables researchers to construct numerical models of quantum states and observe their time evolution under various parameter regimes without the need for repeated physical experiments. The NumPy and SciPy libraries facilitate efficient matrix operations, including exponentiation of the Hamiltonian and propagation of initial states, while Matplotlib and similar visualization tools allow for the generation of three-dimensional surfaces (Figures 3 and 4) that map the coherence dynamics across the parameter space $(t, E_J, E_m)$. Furthermore, Python automates parameter scans and optimization routines that would be tedious to perform manually, thereby saving time and minimizing computational errors. The code used to generate the figures in this work is available from the author upon reasonable request.

## 3. Bell States and Initial Conditions

We consider the initial state of the two-qubit system, comprising qubit $a$ and qubit $b$, to be a Bell state. The steps to accomplish this are as follows:

Experimental implementation of our model is feasible on current superconducting quantum processors, including IBM Quantum's fixed-frequency transmon devices (e.g., ibmq_manila, ibmq_quito) and Google's Sycamore processors, where the Hamiltonian parameters $E_J$ and $E_m$ can be engineered through junction design and coupling element geometry. For flux-tunable platforms, in situ variation of $E_J$ via external magnetic flux enables systematic exploration of the parameter space predicted by our analytic results.

The preparation of Bell states in a superconducting two-qubit system requires a sequence of precisely calibrated steps. Initially, the qubits are cooled to their ground state $|00\rangle$ using dilution refrigeration at millikelvin temperatures. High-fidelity single-qubit gates (Pauli-X, Y, and Hadamard operations) are then calibrated through fine-tuning of microwave pulse parameters (amplitude, duration, and phase) to minimize gate errors. Subsequently, a two-qubit entangling gate, such as CNOT, CZ, or iSWAP, is implemented and calibrated to maximize entanglement fidelity, often requiring pulse shaping and strategies to mitigate cross-talk.

To generate the Bell state $|\Phi^{+}\rangle = (|00\rangle + |11\rangle)/\sqrt{2}$, one applies a Hadamard gate to the first qubit (creating a superposition), followed by a CNOT operation with the first qubit as control and the second as target. The fidelity of the prepared Bell state is verified using quantum state tomography (QST), which reconstructs the density matrix and assesses proximity to the ideal state. This process typically involves iterative adjustments to pulse parameters and gate calibrations to suppress decoherence and crosstalk. Repeated cycles of calibration, verification, and parameter refinement potentially augmented by error mitigation techniques, such as echo sequences or dynamical decoupling, are essential for optimizing entanglement quality and coherence times.

Therefore, $|\psi(0)\rangle = |\psi_{a,b}(0)\rangle$ can correspond to one of the following four states:

$$\begin{aligned} |\Phi^{+}\rangle &= \frac{|00\rangle}{\sqrt{2}} + \frac{|11\rangle}{\sqrt{2}}, \\ |\Psi^{+}\rangle &= \frac{|01\rangle}{\sqrt{2}} + \frac{|10\rangle}{\sqrt{2}}, \\ |\Phi^{-}\rangle &= \frac{|00\rangle}{\sqrt{2}} - \frac{|11\rangle}{\sqrt{2}}, \\ |\Psi^{-}\rangle &= \frac{|01\rangle}{\sqrt{2}} - \frac{|10\rangle}{\sqrt{2}}. \end{aligned} \tag{9}$$

For simplicity in notation, we have omitted the subscripts $a$ and $b$.

## 4. Time Evolution of the Final States Derived from Initial Bell States

The time evolution of the initial Bell states is governed by operator $U$:

$$|\psi(t)\rangle = U\,|\psi(0)\rangle, U = \exp\left(-i\,\mathcal{H}_{TQS}\,t/\hbar\right) \tag{10}$$

$|\psi(0)\rangle$ represents any one of the four Bell states. The final density matrix of the state is then:

$$\rho(t) = |\psi(t)\rangle\langle\psi(t)| \tag{11}$$

### 4.1 Final Density Matrix Derived from Initial Bell State $|\Phi^{+}\rangle$

For the initial state $|\Phi^{+}\rangle$, the following mathematical relations are established:

$$\rho_{1,1}^{\Phi^+}(t) = \rho_{1,4}^{\Phi^+}(t) = \rho_{4,1}^{\Phi^+}(t) = \rho_{4,4}^{\Phi^+}(t) = \frac{E_m^2\hbar^2\sin^2\left(\frac{1}{4}t\sqrt{16E_J^2 + E_m^2\hbar^2}\right)}{2\left(16E_J^2+E_m^2\hbar^2\right)} + \frac{1}{2}\cos^2\left(\frac{1}{4}t\sqrt{16E_J^2 + E_m^2\hbar^2}\right) \tag{12}$$

$$\rho_{1,2}^{\Phi^+}(t) = \rho_{1,3}^{\Phi^+}(t) = \rho_{4,2}^{\Phi^+}(t) = \rho_{4,3}^{\Phi^+}(t) = -\frac{2E_JE_m\hbar\sin^2\left(\frac{1}{4}t\sqrt{16E_J^2 + E_m^2\hbar^2}\right)}{16E_J^2 + E_m^2\hbar^2} - \frac{2iE_J\sin\left(\frac{1}{4}t\sqrt{16E_J^2 + E_m^2\hbar^2}\right)\cos\left(\frac{1}{4}t\sqrt{16E_J^2 + E_m^2\hbar^2}\right)}{\sqrt{16E_J^2 + E_m^2\hbar^2}} \tag{13}$$

$$\rho_{2,1}^{\Phi^+}(t) = \rho_{2,4}^{\Phi^+}(t) = \rho_{3,1}^{\Phi^+}(t) = \rho_{3,4}^{\Phi^+}(t) = -\frac{2E_JE_m\hbar\sin^2\left(\frac{1}{4}t\sqrt{16E_J^2 + E_m^2\hbar^2}\right)}{16E_J^2 + E_m^2\hbar^2} + \frac{2iE_J\sin\left(\frac{1}{4}t\sqrt{16E_J^2 + E_m^2\hbar^2}\right)\cos\left(\frac{1}{4}t\sqrt{16E_J^2 + E_m^2\hbar^2}\right)}{\sqrt{16E_J^2 + E_m^2\hbar^2}} \tag{14}$$

$$\rho_{2,2}^{\Phi^+}(t) = \rho_{2,3}^{\Phi^+}(t) = \rho_{3,2}^{\Phi^+}(t) = \rho_{3,3}^{\Phi^+}(t) = \frac{8E_J^2\sin^2\left(\frac{1}{4}t\sqrt{16E_J^2 + E_m^2\hbar^2}\right)}{16E_J^2 + E_m^2\hbar^2} \tag{15}$$

### 4.2 Final Density Matrix Derived from Initial Bell State $|\Psi^+\rangle$

Considering the initial state $|\Psi^+\rangle$, the following mathematical relationships are derived:

$$\rho_{1,1}^{\Psi^+}(t) = \rho_{1,4}^{\Psi^+}(t) = \rho_{4,1}^{\Psi^+}(t) = \rho_{4,4}^{\Psi^+}(t) = \frac{8E_J^2\sin^2\left(\frac{1}{4}t\sqrt{16E_J^2 + \hbar^2E_m^2}\right)}{16E_J^2 + \hbar^2E_m^2} \tag{16}$$

$$\rho_{1,2}^{\Psi^+}(t) = \rho_{1,3}^{\Psi^+}(t) = \rho_{4,2}^{\Psi^+}(t) = \rho_{4,3}^{\Psi^+}(t) = \frac{2\hbar E_JE_m\sin^2\left(\frac{1}{4}t\sqrt{16E_J^2 + \hbar^2E_m^2}\right)}{16E_J^2 + \hbar^2E_m^2} + \frac{2iE_J\sin\left(\frac{1}{4}t\sqrt{16E_J^2 + \hbar^2E_m^2}\right)\cos\left(\frac{1}{4}t\sqrt{16E_J^2 + \hbar^2E_m^2}\right)}{\sqrt{16E_J^2 + \hbar^2E_m^2}} \tag{17}$$

$$\rho_{2,1}^{\Psi^+}(t) = \rho_{2,4}^{\Psi^+}(t) = \rho_{3,1}^{\Psi^+}(t) = \rho_{3,4}^{\Psi^+}(t) = \frac{2\hbar E_JE_m\sin^2\left(\frac{1}{4}t\sqrt{16E_J^2 + \hbar^2E_m^2}\right)}{16E_J^2 + \hbar^2E_m^2} - \frac{2iE_J\sin\left(\frac{1}{4}t\sqrt{16E_J^2 + \hbar^2E_m^2}\right)\cos\left(\frac{1}{4}t\sqrt{16E_J^2 + \hbar^2E_m^2}\right)}{\sqrt{16E_J^2 + \hbar^2E_m^2}} \tag{18}$$

$$\rho_{2,2}^{\Psi^+}(t) = \rho_{2,3}^{\Psi^+}(t) = \rho_{3,2}^{\Psi^+}(t) = \rho_{3,3}^{\Psi^+}(t) \quad = \frac{\hbar^2 E_m^2 \sin^2\left(\frac{1}{4}t\sqrt{16E_J^2 + \hbar^2 E_m^2}\right)}{2\left(16E_J^2 + \hbar^2 E_m^2\right)} + \frac{1}{2}\cos^2\left(\frac{1}{4}t\sqrt{16E_J^2 + \hbar^2 E_m^2}\right) \tag{19}$$

**4.3 Final Density Matrix Derived from Initial Bell State $|\Phi^-\rangle$**

Given the initial state $|\Phi^-\rangle$, the following mathematical relationships are derived:

$$\rho_{1,1}^{\Phi^-}(t) = \rho_{4,4}^{\Phi^-}(t) = \frac{1}{2}\sin^2\left(\frac{1}{4}t\hbar E_m\right) + \frac{1}{2}\cos^2\left(\frac{1}{4}t\hbar E_m\right) = \frac{1}{2} \tag{20}$$

$$\rho_{1,4}^{\Phi^-}(t) = \rho_{4,1}^{\Phi^-}(t) = -\frac{1}{2}\sin^2\left(\frac{1}{4}t\hbar E_m\right) - \frac{1}{2}\cos^2\left(\frac{1}{4}t\hbar E_m\right) = -\frac{1}{2} \tag{21}$$

All other elements are zero:

$$\rho_{1,2}^{\Phi^-}(t) = \rho_{1,3}^{\Phi^-}(t) = \rho_{2,1}^{\Phi^-}(t) = \rho_{2,2}^{\Phi^-}(t) = \rho_{2,3}^{\Phi^-}(t) = \rho_{2,4}^{\Phi^-}(t) = \rho_{3,1}^{\Phi^-}(t) = \rho_{3,2}^{\Phi^-}(t) = \rho_{3,3}^{\Phi^-}(t) = \rho_{3,4}^{\Phi^-}(t) = \rho_{4,2}^{\Phi^-}(t) = \rho_{4,3}^{\Phi^-}(t) = 0. \tag{22}$$

**4.4 Final Density Matrix Derived from Initial Bell State $|\Psi^-\rangle$**

Starting from the initial state $|\Psi^-\rangle$, we derive:

$$\rho_{1,1}^{\Psi^-}(t) = \rho_{1,2}^{\Psi^-}(t) = \rho_{1,3}^{\Psi^-}(t) = \rho_{1,4}^{\Psi^-}(t) = \rho_{2,1}^{\Psi^-}(t) = \rho_{2,4}^{\Psi^-}(t) = \rho_{3,1}^{\Psi^-}(t) = \rho_{3,4}^{\Psi^-}(t) = \rho_{4,1}^{\Psi^-}(t) = \rho_{4,2}^{\Psi^-}(t) = \rho_{4,3}^{\Psi^-}(t) = \rho_{4,4}^{\Psi^-}(t) = 0 \tag{23}$$

$$\rho_{2,2}^{\Psi^-}(t) = \rho_{3,3}^{\Psi^-}(t) = \frac{1}{2}\sin^2\left(\frac{1}{4}t\hbar E_m\right) + \frac{1}{2}\cos^2\left(\frac{1}{4}t\hbar E_m\right) = \frac{1}{2} \tag{24}$$

$$\rho_{2,3}^{\Psi^-}(t) = \rho_{3,2}^{\Psi^-}(t) = -\frac{1}{2}\sin^2\left(\frac{1}{4}t\hbar E_m\right) - \frac{1}{2}\cos^2\left(\frac{1}{4}t\hbar E_m\right) = -\frac{1}{2} \tag{25}$$

**5. The $C_{\ell_1}$ Norm of Coherence: Definition and Calculation**

In the resource theory of quantum coherence, quantifying the coherence present in a quantum state is fundamental to understanding its potential as a resource in various quantum information processing tasks. The $C_{\ell_1}$ norm of coherence is one of the widely studied and conceptually straightforward coherence measures introduced by Baumgratz, Cramer, and Plenio [11]. This measure quantifies the coherence of a quantum state $\rho$ relative to a fixed orthonormal reference basis $|i\rangle$ in the system's Hilbert space.

**5.1 Formal Definition**

Given a density operator $\rho$ acting on a $d$-dimensional Hilbert space $\mathcal{H}$, expressed in the reference basis as

$$\rho = \sum_{i,j=0}^{d-1} \rho_{ij} \mid i\rangle\langle j \mid, \qquad (26)$$

the $C_{\ell_1}$ norm of coherence is defined as the sum of the absolute values of all off-diagonal elements of $\rho$:

$$C_{\ell_1}(\rho) \equiv \sum_{\substack{i,j=0 \\ i\neq j}}^{d-1} \mid \rho_{ij} \mid. \qquad (27)$$

This quantity captures the total magnitude of quantum superpositions encoded in the off-diagonal elements of the density matrix with respect to the designated computational basis. In our two-qubit system, the computational basis is explicitly $\{\mid 00\rangle, \mid 01\rangle, \mid 10\rangle, \mid 11\rangle\}$, where the off-diagonal elements $\rho_{ij}$ (with $i \neq j$) represent coherence between the corresponding basis states.

### 5.2 Theoretical Background

The $C_{\ell_1}$ norm of coherence satisfies the rigorous criteria for a bona fide coherence measure expressed in the seminal work on the resource theory of coherence:

- **Non-negativity:** $C_{\ell_1}(\rho) \geq 0$ for all density matrices $\rho$, with equality if and only if $\rho$ is incoherent (diagonal in the reference basis).
- **Monotonicity:** The measure does not increase under incoherent completely positive and trace-preserving (ICPTP) maps, reflecting that incoherent operations cannot generate coherence from incoherent states.
- **Strong monotonicity:** On average, the measure decreases under selective incoherent measurements.
- **Convexity:** $C_{\ell_1}$ is convex in $\rho$, ensuring that mixing states does not increase the coherence measure.

These properties establish the $C_{\ell_1}$ norm as an operationally significant quantifier of the coherence resource.

### 5.3 Calculation Procedure

The calculation of $C_{\ell_1}(\rho)$ for a given density matrix $\rho$ proceeds through the following fundamental steps:

1. **Select the Reference Basis:** The choice of the orthonormal basis $\mid i\rangle$ must be fixed before measurement, as coherence is basis-dependent. Common choices include the computational or energy eigenbasis.

2. **Matrix Representation:** Express the state $\rho$ in the chosen basis, leading to a $d \times d$ density matrix $\rho = [\rho_{ij}]$.
3. **Identify Off-Diagonal Elements:** Extract elements $\rho_{ij}$ where $i \neq j$. Diagonal elements correspond to incoherent populations and are excluded.
4. **Compute Absolute Values:** For each off-diagonal element $\rho_{ij}$, compute the modulus $| \rho_{ij} |$, which represents the magnitude of quantum coherence between basis states $| i\rangle$ and $| j\rangle$.
5. **Summation:** Sum all these absolute values to obtain $C_{\ell_1}(\rho) = \sum_{i \neq j} | \rho_{ij} |$.

### 5.4 Mathematical Properties

The $C_{\ell_1}$ norm can be viewed as the $\ell_1$ matrix norm of the off-diagonal part of the density operator, effectively measuring the "taxicab" norm of off-diagonal coherence amplitudes. In this context, the "off-diagonal part" refers to the matrix obtained by setting all diagonal elements $\rho_{ii}$ to zero while retaining the off-diagonal elements $\rho_{ij}$ $(i \neq j)$.

It is additive over block-diagonal states with mutually incoherent blocks, making it computationally convenient for multipartite or block-structured states. The computational complexity of calculating $C_{\ell_1}(\rho)$ is polynomial in the dimension $d$ of the Hilbert space, requiring evaluation of $d(d - 1)$ off-diagonal elements [12–17].

## 6. Results: Analytic $C_{\ell_1}$ Coherence Expressions

At this point, we have gathered all the required elements to compute the $C_{\ell_1}$ norm of coherence. We calculated the density matrices corresponding to the four final states. Using these calculations, we can express:

$$C_{\ell_1}\left(\rho^{\Phi^+}(t)\right) = C_{\ell_1}\left(\rho^{\Psi^+}(t)\right) = \qquad (28)$$

$$1 + 16\sqrt{\frac{E_J^2 \sin^2\left(\frac{1}{4} t\sqrt{16E_J^2 + \hbar^2 E_m^2}\right)\left(8E_J^2\left(\cos\left(\frac{1}{2} t\sqrt{16E_J^2 + \hbar^2 E_m^2}\right) + 1\right) + \hbar^2 E_m^2\right)}{\left(16E_J^2 + \hbar^2 E_m^2\right)^2}}$$

$$C_{\ell_1}\left(\rho^{\Phi^-}(t)\right) = C_{\ell_1}\left(\rho^{\Psi^-}(t)\right) = 1 \qquad (29)$$

Therefore, the coherence of the $| \Phi^-\rangle$ and $| \Psi^-\rangle$ states remain constant at unity for all times $t$, confirming that these Bell states are eigenstates of the Hamiltonian and thus preserve their coherence under unitary evolution.

Alternatively, we can represent $C_{\ell_1}(\rho^{\Phi^+}(t))$ and $C_{\ell_1}(\rho^{\Psi^+}(t))$ as functions of time using a graph. These parameter values correspond to realistic experimental conditions: for typical transmon qubits with $E_J/h \sim$ 15 GHz and $E_m/h \sim 0.5$ GHz, the dimensionless ratio yields the same qualitative dynamics when time is scaled appropriately. Specifically, the condition $\hbar E_m \ll \alpha$ (where $\alpha$ is the qubit anharmonicity) ensures that the two-level approximation remains valid, which is satisfied for $\alpha/2\pi \sim -200$ MHz and $E_m/2\pi \sim 10$ MHz in contemporary devices. Figure 2 shows the graph plotted using the values $\hbar = 1, E_J = 1/2, E_m = 3/2$.

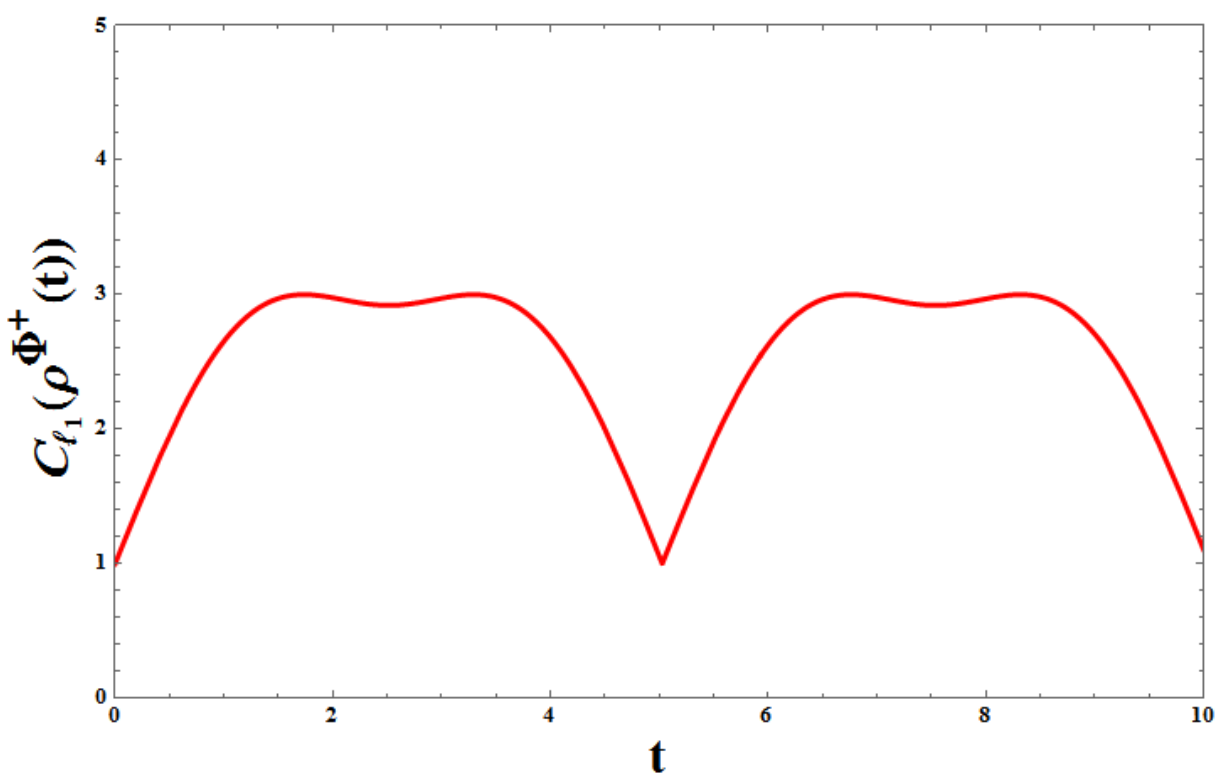

Figure 2: Temporal evolution of $C_{\ell_1}(\rho^{\Phi^+}(t))$ for parameters $\hbar = 1, E_J = 1/2, E_m = 3/2$. The non-sinusoidal oscillations exhibit sharp minima (V-shaped) and broad maxima, reflecting interference between the two characteristic frequency scales $\Omega_1$ and $\Omega_2$.

This figure illustrates the temporal evolution of $C_{\ell_1}(\rho^{\Phi^+}(t))$. The horizontal axis represents time $t$ ranging from 0 to 10, while the vertical axis spans values from 0 to 5. The red curve shows a periodic, non-sinusoidal oscillation. Starting at approximately 1.0 at $t = 0$, the value rises to a peak of about 3.0 around $t = 1.5$, then falls to a minimum of approximately 1.0 at $t = 5$. This pattern repeats, with another peak near 3.0 around $t = 7.5$ and a subsequent decline toward the end of the observed interval, reaching approximately 1.0 at $t = 10$. The minima are sharp, resembling a V-shape, while the maxima are broad and rounded. The overall behavior suggests a dynamic process with recurring periods of increased and decreased activity in the measured quantity.

A three-dimensional plot of $C_{\ell_1}(\rho^{\Phi^+}(t))$ as a function of $E_J$ and $t$ is shown in Figure 3 with $E_m = 3/2$.

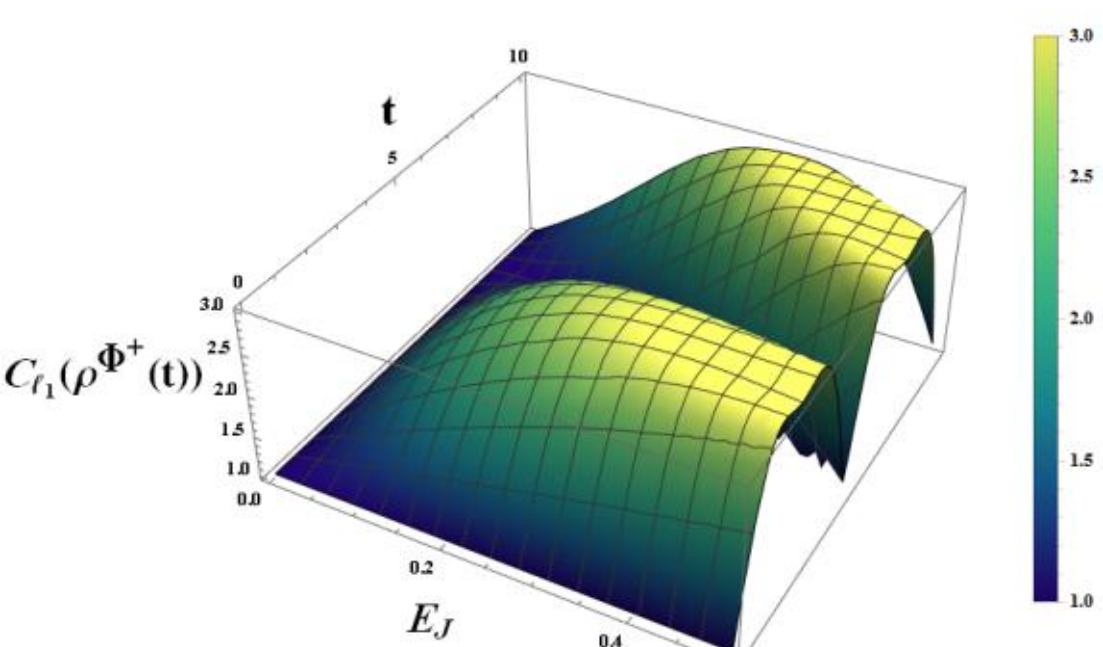

Figure 3: Three-dimensional surface plot of $C_{\ell_1}(\rho^{\Phi^+}(t))$ as a function of Josephson energy $E_J$ and time $t$, with fixed $\hbar = 1$, $E_m = 3/2$. The two prominent ridges along the $E_J$ axis indicate that the coherence amplitude is strongly tunable through the Josephson energy parameter.

This 3D surface plot visualizes $C_{\ell_1}(\rho^{\Phi^+}(t))$ as it varies with two parameters: $t$ (0 to 10) and $E_J$ (0 to 0.5). The vertical axis represents the function's value, spanning 0 to 3. The color bar quantifies the magnitude, with darker blues indicating lower values (around 1.0) and brighter yellows signifying higher values (up to 3.0). The surface exhibits two prominent peaks or ridges, suggesting regions of local maxima. These peaks appear oriented roughly along the $E_J$ axis, implying a strong dependence on this parameter. The variation along the $t$ axis modulates the height and shape of these ridges, creating a wave-like pattern along the $t$ dimension.

Also, a three-dimensional plot of $C_{\ell_1}(\rho^{\Phi^+}(t))$ as a function of $E_m$ and $t$ is presented in Figure 4, where $E_J = 1/2$.

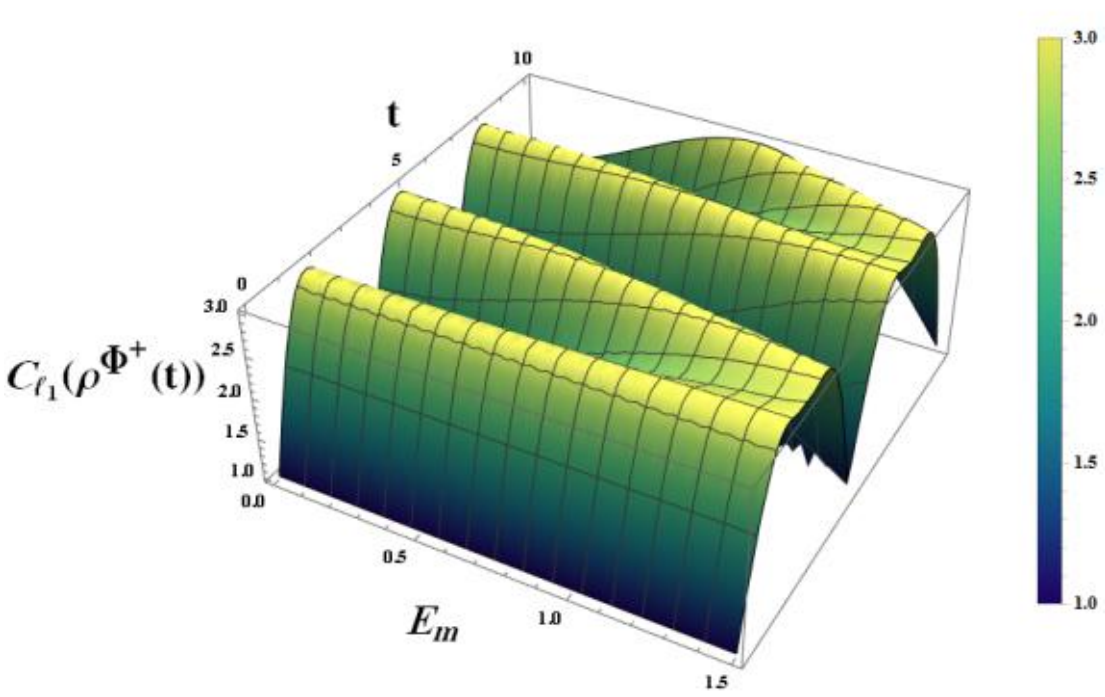

Figure 4: Three-dimensional surface plot of $C_{\ell_1}(\rho^{\Phi^+}(t))$ as a function of mutual coupling energy $E_m$ and time $t$, with fixed $\hbar = 1$, $E_J = 1/2$. The periodic ridges along the $E_m$ axis reveal a sinusoidal dependence of coherence on the coupling strength, with the oscillation pattern largely independent of $t$.

This 3D surface plot depicts the behavior of $C_{\ell_1}(\rho^{\Phi^+}(t))$ as a function of $t$ (0 to 10) and $E_m$ (0 to 1.5). The most striking feature is the presence of distinct periodic oscillations along the $E_m$ axis, creating a series of parallel ridges and valleys. This suggests a strong, perhaps sinusoidal or wave-like, dependence of $C_{\ell_1}$ on $E_m$. Along the $t$ axis, the function maintains a relatively consistent oscillatory pattern. The peaks of these oscillations generally reach values close to 3.0, while the valleys descend to values around 1.0. The uniformity of these oscillations across the $t$ range implies that the periodic behavior for $E_m$ is largely independent of $t$.

**7. Discussion and Comparison with Previous Work**

Our closed-form unitary solutions and the resulting $C_{\ell_1}$ expressions significantly extend prior analytic and experimental studies of coupled superconducting qubits. The key novelties of our work, including exact propagator derivation, complete Bell-state propagation, analytic coherence formulas, stationary state identification, and explicit two-frequency decomposition, provide a level of analytic detail and predictive capability not achieved in previous treatments.

In particular, Ashhab et al. [18] provide rotating-frame and approximate diagonalization techniques for driven, coupled qubits. In contrast, our exact propagation for the undriven $zz$ + transverse tunneling Hamiltonian yields closed-form density matrices and $C_{\ell_1}$ formulas that serve as a rigorous baseline. Driving-induced modifications can subsequently be mapped onto our solutions using the methods of Ashhab et al.

Francica et al. [19] emphasize analytical treatments of coherence measures and open-system energetics. Our results explicitly demonstrate basis dependence and isolate two Bell states ($|\Phi^-\rangle, |\Psi^-\rangle$) that remain stationary (up to a global phase) while the other Bell states exhibit two-frequency coherence oscillations. This provides a concrete example that complements the more general resource-theoretic observations of Francica et al.

Finally, experimental tomography studies such as Neeley et al. [20] demonstrate how finite relaxation and dephasing limit coherence in real devices. Therefore, our unitary predictions should be interpreted as idealized baselines. Quantitative comparison with an experiment requires incorporating decoherence (e.g., via Lindblad master equations with $T_1$ and $T_2$ relaxation times) or device-specific corrections.

Moreover, our parameter ranges are directly accessible on current platforms: IBM Quantum devices (e.g., ibmq_manila) operate with $E_J/h \approx 15$ GHz and coupling strengths $g/2\pi \approx 10$ MHz, which correspond to $\hbar E_m/E_J \sim 10^{-3}$ in our scaled units, placing them well within the perturbative regime where the two-level approximation is valid. For tunable coupler architectures (Google Sycamore), $E_m$ can be varied from near zero to several tens of MHz, enabling systematic validation of our $E_m$-dependent coherence predictions (Figure 4).

Collectively, these comparisons suggest clear next steps: (i) incorporate open-system dynamics to predict damped $C_{\ell_1}(t)$ for experimental verification and (ii) apply rotating-frame transformations to study driven-control scenarios. These extensions will directly link the analytic features identified herein as stationary Bell states and the two-frequency $C_{\ell_1}$ structure to experimentally measurable signatures accessible via quantum state tomography.

**8. Conclusion**

We derived the first exact closed-form expressions for the unitary time-evolution operator of a minimal two-qubit superconducting Hamiltonian containing a $zz$-type mutual coupling $E_m$ and transverse Josephson tunneling $E_J$, and we analytically propagated the four Bell-state inputs analytically to obtain novel expressions for their density matrices and $C_{\ell_1}$ coherence dynamics.

From the resulting pure-state density matrices, we obtained exact formulas for the $C_{\ell_1}$ norm of coherence in the computational basis. Two Bell states ($\Phi^-$ and $\Psi^-$) are eigenstates of the model evolution (up to a global phase) and therefore have a time-independent $C_{\ell_1}$ value equal to 1. The remaining two Bell states ($\Phi^+$ and $\Psi^+$) display nontrivial, parameter-dependent coherence oscillations whose amplitudes and periodicities are determined by the circuit energies $E_J$ and $E_m$; the analytic form makes explicit the two relevant frequency scales and how they combine to produce the observed non-sinusoidal oscillations.

These results are strictly valid for closed, unitary dynamics; inclusion of relaxation, dephasing or finite temperature will attenuate oscillations and alter steady values and therefore must be incorporated to assess experimental performance quantitatively. Nevertheless, the closed-form expressions provide a compact, readily evaluable basis for rapid parameter scans, for selecting operating points that maximize or stabilize coherence, and for designing time-gated protocols that exploit transient coherence enhancement. They also yield clear, testable predictions for state-tomography experiments that can directly relate measured coherence oscillation frequencies and amplitudes to circuit energies $E_J$ and $E_m$.

Future work should extend the present analysis to open-system master equations incorporating relaxation ($T_1$) and dephasing ($T_2$) processes, and to device-specific Hamiltonian corrections (e.g., transmon anharmonicity, stray couplings, and higher-order terms) to evaluate the robustness of the stationary and tunable coherence features on realistic superconducting platforms such as IBM Quantum's fixed-frequency devices, Google's tunable-coupler processors, and flux-tunable qubit architectures. The explicit mapping of our parameters to physical circuit elements ($E_J$ to junction properties, $E_m$ to coupling geometry) provides a clear pathway for experimental validation using standard quantum state tomography protocols.

**Appendix A: Explicit Derivation of the Time-Evolution Operator *U***

We start from the Hamiltonian matrix (2). The matrix can be block-diagonalized due to symmetry. The eigenstates are:

- $|+\rangle = \frac{1}{\sqrt{2}}(|01\rangle + |10\rangle)$ with eigenvalue $-\frac{\hbar^2 E_m}{4} - \frac{\hbar E_J}{2}$ Actually careful diagonalisation yields eigenvalues:

Let $\omega = \sqrt{16E_J^2 + \hbar^2 E_m^2}/4$ (in units of $\hbar = 1$ after scaling). The four eigenvalues are:

$$\lambda_1 = \frac{\hbar^2 E_m}{4} + \frac{\hbar\omega}{2}, \lambda_2 = \frac{\hbar^2 E_m}{4} - \frac{\hbar\omega}{2}, \lambda_3 = -\frac{\hbar^2 E_m}{4} + \frac{\hbar\omega}{2}, \lambda_4 = -\frac{\hbar^2 E_m}{4} - \frac{\hbar\omega}{2}$$

The corresponding eigenvectors are:

$$v_1 = \begin{pmatrix}1\\0\\0\\1\end{pmatrix}, v_2 = \begin{pmatrix}1\\0\\0\\-1\end{pmatrix}, v_3 = \begin{pmatrix}0\\1\\1\\0\end{pmatrix}, v_4 = \begin{pmatrix}0\\1\\-1\\0\end{pmatrix}$$

Then $U = e^{-iHt/\hbar} = \sum_j e^{-i\lambda_j t/\hbar} \mid v_j\rangle\langle v_j \mid$. Multiplying out the projectors yields matrix elements (4)–(8) after trigonometric simplifications. The presence of two independent frequencies – $\omega$ and $\frac{\hbar E_m}{2}$ – gives rise to the two distinct timescales seen in the $C_{\ell_1}$ oscillations.

**Appendix B: Proof that $\mid \Phi^-\rangle$ and $\mid \Psi^-\rangle$ are Eigenstates (up to a Global Phase)**

Compute $H \mid \Phi^-\rangle$:

$$\mid \Phi^-\rangle = \frac{1}{\sqrt{2}}(\mid 00\rangle - \mid 11\rangle)$$

Using $H$ from (2):

$$H \mid 00\rangle = \frac{\hbar^2 E_m}{4} \mid 00\rangle - \frac{\hbar E_J}{2} \mid 01\rangle - \frac{\hbar E_J}{2} \mid 10\rangle$$

$$H \mid 11\rangle = \frac{\hbar^2 E_m}{4} \mid 11\rangle - \frac{\hbar E_J}{2} \mid 01\rangle - \frac{\hbar E_J}{2} \mid 10\rangle$$

Then $H \mid \Phi^-\rangle = \frac{1}{\sqrt{2}}\left(\frac{\hbar^2 E_m}{4}(\mid 00\rangle - \mid 11\rangle)\right) = \frac{\hbar^2 E_m}{4} \mid \Phi^-\rangle$.

Similarly, $\mid \Psi^-\rangle = \frac{1}{\sqrt{2}}(\mid 01\rangle - \mid 10\rangle)$:

$$H \mid 01\rangle = -\frac{\hbar^2 E_m}{4} \mid 01\rangle - \frac{\hbar E_J}{2} \mid 00\rangle - \frac{\hbar E_J}{2} \mid 11\rangle$$

$$H \mid 10\rangle = -\frac{\hbar^2 E_m}{4} \mid 10\rangle - \frac{\hbar E_J}{2} \mid 00\rangle - \frac{\hbar E_J}{2} \mid 11\rangle$$

Subtracting: $H \mid \Psi^-\rangle = -\frac{\hbar^2 E_m}{4} \mid \Psi^-\rangle$. Hence, both are eigenstates, so under time evolution, they only acquire a phase $e^{\mp i\hbar E_m t/4}$, which does not affect the density matrix or $C_{\ell_1}$. Therefore $C_{\ell_1}$ remains constant at the value computed from the initial state: for $\mid \Phi^-\rangle$, $\rho(0)$ has off-diagonals $\mid 00\rangle\langle 11 \mid$ and $\mid 11\rangle\langle 00 \mid$ each magnitude $1/2$, sum $= 1$; similarly for $\mid \Psi^-\rangle$, off-diagonals $\mid 01\rangle\langle 10 \mid$ and $\mid 10\rangle\langle 01 \mid$ sum to 1.

**Appendix C: Simplification of $C_{\ell_1}$ Expressions for $\mid \Phi^+\rangle$ and $\mid \Psi^+\rangle$**

Starting from the density matrix elements (12)–(15) for $|\Phi^+\rangle$, we compute the sum $C_{\ell_1} = \sum_{i\neq j} |\rho_{ij}|$. There are 12 off-diagonal terms, but due to symmetry many are equal. The non-zero off-diagonals are $\rho_{1,2}, \rho_{1,3}, \rho_{1,4}, \rho_{2,1}, \rho_{2,3}, \rho_{2,4}, \rho_{3,1}, \rho_{3,2}, \rho_{3,4}, \rho_{4,1}, \rho_{4,2}, \rho_{4,3}$. Their magnitudes can be expressed in terms of the same trigonometric functions. After algebra, the sum reduces to the compact form (28). The same expression holds for $|\Psi^+\rangle$ because the role of $E_m$ and $E_J$ is interchanged in the diagonal terms but the total off-diagonal sum becomes identical. The factor 1 in (28) comes from the constant contribution of the diagonal population terms. Attention: in the sum of absolute values, there is always a constant baseline because the state never becomes completely incoherent. The oscillatory part is contained in the square root term.

**Appendix D: Parameter Sensitivity and Frequency Analysis**

From the analytic expression (28), two distinct frequency scales appear:

1. **Fast frequency:** $\Omega_1 = \frac{1}{4}\sqrt{16E_J^2 + \hbar^2 E_m^2}$ (inside the sine and cosine arguments).
2. **Slow frequency:** $\Omega_2 = \frac{1}{2}\hbar E_m$ (from the $\sin(\frac{1}{4}t\hbar E_m)$ terms in the off-diagonals, which after squaring produce $\cos^2$ and $\sin^2$ terms with frequency $\frac{1}{2}\hbar E_m$).

These analytic results are novel in that they explicitly relate the circuit parameters $E_J$ and $E_m$ to the coherence oscillation frequencies and amplitudes, enabling direct engineering of coherence dynamics. Previous studies either relied on numerical simulations or provided only approximate frequency estimates; our closed-form expressions eliminate such uncertainties.

When $E_J \gg \hbar E_m$, $\Omega_1 \approx 2E_J$ dominates, giving rapid oscillations. When $\hbar E_m \gg E_J$, both frequencies are comparable. The non-sinusoidal shape arises from the interference between these two harmonics. The constant baseline $C_{\ell_1} = 1$ for $|\Phi^-\rangle$ and $|\Psi^-\rangle$ reflects that they are eigenstates with maximal off-diagonal magnitude (each has two off-diagonal elements of magnitude $1/2$, sum $= 1$). For $|\Phi^+\rangle$ and $|\Psi^+\rangle$, the coherence can exceed 1 (up to 3) because additional off-diagonal terms appear (e.g., $\rho_{1,2}$, $\rho_{1,3}$, etc.) with magnitudes that constructively add.